\documentclass[manuscript]{aastex631}

\usepackage{wrapfig}
\usepackage{tablefootnote}

\shorttitle{Exploring Cen A with \textit{INTEGRAL}}
\shortauthors{Rodi et al.}
\graphicspath{{./}{figures/}}

\begin{document}

\title{Centaurus A: Exploring the Nature of the Hard X-ray/Soft Gamma-ray Emission with \textit{INTEGRAL}}

\author[0000-0003-2126-5908]{James Rodi}
\affiliation{INAF - Istituto di Astrofisica e Planetologia Spaziali; via Fosso del Cavaliere 100; 00133 Roma, Italy}

\author[0000-0001-9932-3288]{E. Jourdain}
\affiliation{Universit\'e de Toulouse; UPS-OMP; IRAP;  Toulouse, France}
\affiliation{CNRS; IRAP; 9 Av. Colonel Roche, BP 44346, F-31028 Toulouse cedex 4, France}

\author[0000-0003-0681-9984]{M. Molina}
\affiliation{Istituto di Astrofisica Spaziale e Fisica Cosmica di Milano; Milan, Italy}

\author[0000-0002-1757-9560]{J. P. Roques}
\affiliation{Universit\'e de Toulouse; UPS-OMP; IRAP;  Toulouse, France}
\affiliation{CNRS; IRAP; 9 Av. Colonel Roche, BP 44346, F-31028 Toulouse cedex 4, France}



\begin{abstract}

The question of the origin of the hard X-ray/soft gamma-ray emission in Centaurus A (Cen A) persists despite decades of observations.  Results from X-ray instruments suggest a jet origin since the implied electron temperature (\(kT_e\)) would cause pair production runaway in the corona.  In contrast, instruments sensitive to soft gamma-rays report electron temperatures indicating a corona origin may be possible.  In this context, we analyzed archival \textit{INTEGRAL}/IBIS-ISGRI and SPI data and observations from a 2022 Cen A monitoring program.  Our analysis did not find any spectral variability.  Thus we combined all observations for long-term average spectra, which were fit with a \textit{NuSTAR} observation to study the 3.5 keV \(-\) 2.2 MeV spectrum.  Spectral fits using a \texttt{CompTT} model found \(kT_e \sim 550\) keV, near pair-production runaway.  The spectrum was also well described by a log-parabola to model synchrotron self-Compton emission from the jet.  Additionally, a spectral fit with the 12-year catalog \textit{Fermi}/LAT spectrum using a log-parabola can explain the data up to \( \sim 3 \) GeV.  Above \( \sim 3\) GeV, a power-law excess is present, which has been previously reported in LAT/H.~E.~S.~S. analysis.  However, including a corona spectral component can also describe the data well.  In this scenario, the hard X-rays/soft gamma-rays are due the corona and the MeV to GeV emission is due to the jet.



\end{abstract}



\section{Introduction} \label{sec:intro}

As one of the brightest radio-loud active galactic nuclei (AGNs) in the hard X-ray sky, Centaurus A (Cen A, \(z = 0.001826\)) has been observed by numerous missions since its first detection in 1969 \citep{bowyer1970}.  Results have shown that the spectral shape does not vary with flux \citep{jourdain1993,rothschild2006,beckmann2011,rothschild2011,burke2014}.  Consequently, one would expect reported spectral parameters for Cen A to be similar across time and instruments.  Yet that has not been the case as there has been a lack of consistency in the values, which has made interpreting the origin of the hard X-ray/soft gamma-ray emission difficult.  


The presence of spectral curvature and where it occurs have implications on the emission process.  Often, the hard X-ray continuum in AGNs is interpreted as thermal inverse Comptonization of UV seed photons from the accretion disk \citep{haardt1993,merloni2003}.  The spectra typically have emission up to a few 100 keV \citep{molina2013}, and the spectral cutoff energy is related to the temperature of electrons in the corona.  In the corona, photon-photon collisions can create electron-positron pairs when the sum of the photon energies is above \( \sim 1\) MeV \citep{svensson1984,stern1995,fabian2015}.  Thus if the electron temperature is too high, the production of electron-positron pairs can become a runaway process that reduces the corona temperature.  Additional pairs form and gain energy from the corona, which lowers the amount of energy per particle \citep{fabian2015,fabian2017}.

The maximum temperature is related to the geometry of the source, which is typically given by the compactness parameter, \( \ell\), which is the ratio of the source luminosity and its size \citep{guilbert1983}.  The limits on electron temperature range from \( \sim 500\) keV in the case of very low compactness and \( \sim 70\) keV in the case of very high compactness \citep{fabian2017}.  For Cen A, \( \ell \sim 10\) \citep{fabian2017}, suggesting a maximum corona temperature of a few 100 keV.

Such disagreements over the curvature in Cen A can be seen comparing results from the past \( \sim 30\) years with the main disagreement over the location of spectral curvature.   Multiple analyses have found spectra well fit with a power-law model with similar spectral indexes.  \emph{GRANAT}/SIGMA observations in \(1990-1991\) found a power-law spectrum with a slope of \(1.9 \pm 0.3\) (\(35-200\) keV) with only a weak indication of a spectral break \citep{jourdain1993}.  Later observations from \emph{RXTE} and \emph{INTEGRAL} in the \(3-240\) keV energy range found a power-law spectrum with a slope of \(1.83 \pm 0.01\) and no clear indication of a spectral break \citep{rothschild2006}.  Also, Rothschild et al. (2011) analysis using approximately 12 years of \emph{RXTE} data supported earlier findings with a slope of \(1.822 \pm 0.004\) and no evidence of curvature, constraining the cutoff to \( > 2\) MeV \citep{rothschild2011}.  More recently, \textit{XMM-Newton} and \textit{NuSTAR} observations found an index of 1.81 with no need for a cutoff \( < 1\) MeV (Furst et al. 2016).  These high-energy spectral cutoff energies are consistent with the hard X-rays as part of the inverse Compton hump of the spectral energy distribution (SED), which \cite{chiaberge2001} interpreted as synchrotron self-Compton (SSC) emission.  However, they do find the data is consistent with a thermal Comptonization model with an electron temperature \( \sim 100-300\) keV.  All of these proposed spectral turnovers in the above results are well outside the energy ranges of the considered spectra and thus are not able to constrain the spectral break.

Unfortunately, observations at soft gamma-ray energies have not fared better at determining where the spectral break is.  Joint spectral fits from the OSSE and COMPTEL instruments (50 keV \(-\) 30 MeV) found a spectral break at \( \sim 150\) keV and a low energy power-law slope of 1.74 and high energy power-law slope of 2.0 \citep{steinle1998}.  \emph{INTEGRAL} observations covering \(3 - 1000\) keV analyzed by \cite{beckmann2011} can be fit with a cutoff power-law with a photon index of  \( \sim 1.7 \textrm{ and E}_c \sim 400\) keV and excluding cutoff energies below 300 keV and above 700 keV at \( 3 \sigma\) so significantly different from the spectral cutoff above \( 1-2\) MeV predicted from lower energy instruments.  Analysis of \emph{INTEGRAL}/SPI with \emph{Chandra} in the \(2 - 500\) keV energy range also found spectral curvature.  Their analysis, which included a reflection component, found a cutoff energy above 700 keV with a confidence of \(> 95 \%\) \citep{burke2014}.

\cite{skibo1994} proposed an interesting alternative where the emission is the result of beamed photons Compton scattered into our line-of-sight.  Using OSSE data in the \(50 - 1000\) keV range, they found a photon index of \(1.68 \pm 0.03\) and an angle of \(\sim 61^{\circ}\) between the jet and our line-of-sight assuming the material scattering the beamed emission is non-relativistic.

In this work, we report on additional \textit{INTEGRAL} observations of Cen A spanning nearly 20 years, which allows us to explore the long-term temporal variability and study the soft gamma-ray spectrum to investigate the possible high-energy cutoff, which provides key information on the emission process at these energies.

\section{Instruments and Observations} \label{sec:inst}

\subsection{INTEGRAL}

The \textit{INTEGRAL} satellite was launched in October 2002  from Baikonur, Kazachstan with an eccentric orbit \citep{jensen2003} that has a period of \( \sim 2.5 - 3\) days.  We analyzed data from both SPI (SPectrometer on-board INTEGRAL \citep{vedrenne2003}) and IBIS/ISGRI (Imager on Board INTEGRAL/INTEGRAL Soft Gamma-Ray Imager \citep{ubertini2003}).  SPI spans the \(20\) keV \(-\) 8 MeV energy range with an energy resolution of \( 2-8 \) keV \citep{roques2003}, and ISGRI covers energies from \(15 - 1000\) keV.  

The SPI data were analyzed using the SPI Data Analysis Interface (SPIDAI)\footnote{Publicly available interface developed at IRAP to analyze SPI data. Available by contacting:  spidai@irap.omp.eu. See description in \citet{burke2014}} to generate light curves and spectra, which consist of 39 channels over the 20 keV \( - \) 2.2 MeV range.  The data above 400 keV were corrected for Pulse Shape Discriminator efficiency following the method outlined in \cite{roques2019}.  The ISGRI data were analyzed with the Offline Scientific Analysis 11.2 (OSA 11.2)\footnote{https://www.isdc.unige.ch/integral/analysis\#Software} to produce light curves and spectra, which span the \(30 - 630 \) keV range in 43 channels with a 1\% systematic error included.

The observations used in this work span from revolution 48 to 2590 (MJD 52705 \(-\) 59948 or 2003-03-07 to 2023-01-04) when the source was within \(10^{\circ}\) from the \textit{INTEGRAL} pointing direction and incorporates archival data and data from our current observing campaign during \textit{INTEGRAL}'s AO-19 observation cycle.  Each observation or science window (scw) is \( \sim 1800 - 3600\) s.  Observations contaminated by solar activity and Earth's radiation belt have been removed.  Thus the total exposure time is 5.3 Ms for SPI and 4.6 Ms for ISGRI.

\subsection{NuSTAR}
\textit{NuSTAR} data (from both focal plane detectors, FPMA and FPMB)
were reduced using the \texttt{nustardas\_04May21\_v2.1.1} and \texttt{CALDB} version
20220118. In this work, we reduced observation 60001081002, taken Aug. 6, 2013,
with a cleaned exposure of $\sim$51.3 ksec; spectral extraction and the subsequent production of
response and ancillary files were performed using the
\texttt{nuproducts} task with an extraction radius of $\sim$136$^{\prime\prime}$; 
to maximise the signal-to-noise ratio; background spectra were extracted from 
circular regions of 70$^{\prime\prime}$ radius 
in source-free areas of the detectors. 

 \begin{figure}[h!]
  \begin{center}
  \includegraphics[scale=0.7, angle=180,trim = 5mm 5mm 20mm 0mm, clip]{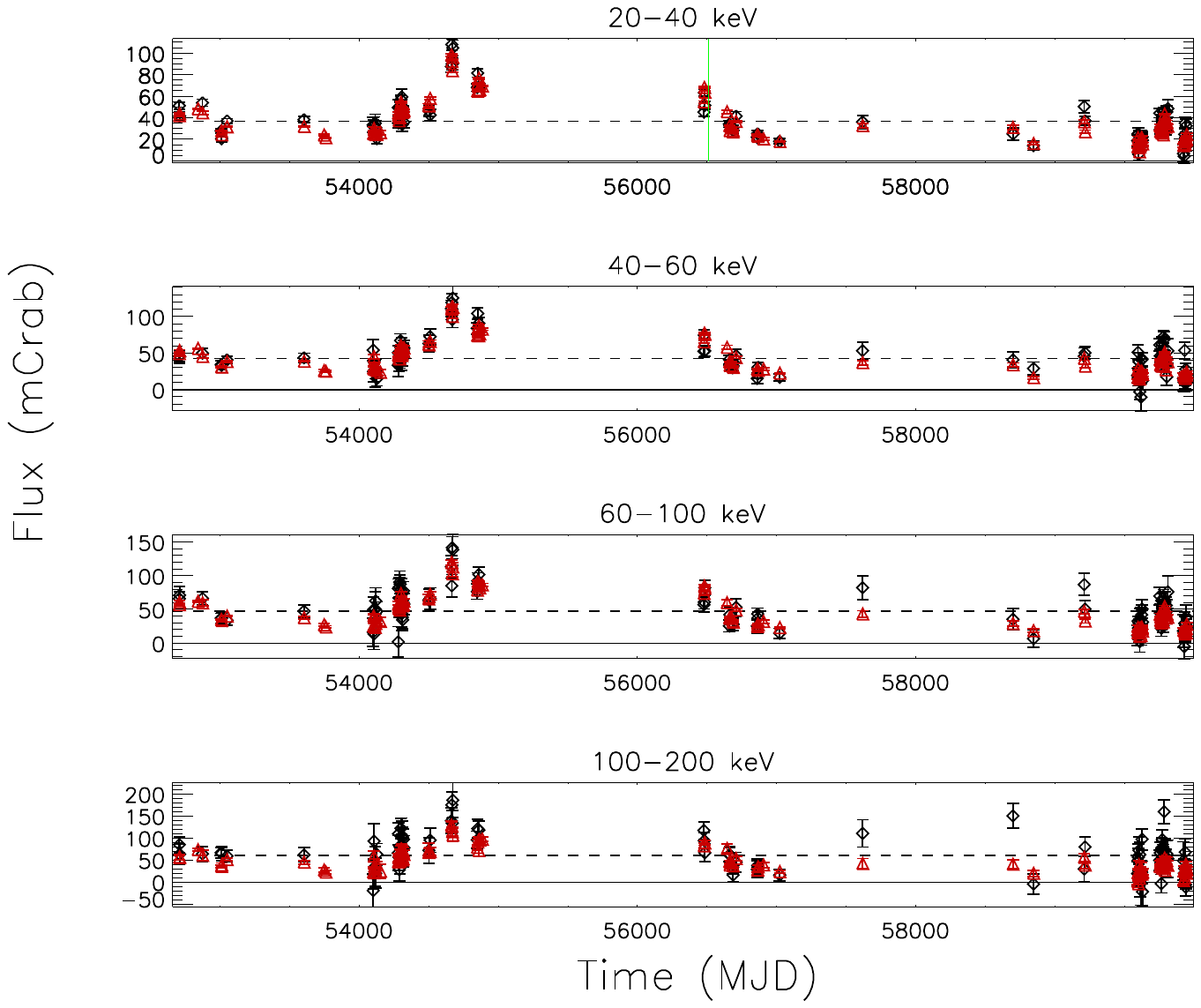}
    \end{center}
  \caption{Cen A SPI (black diamonds) and ISGRI (red triangles) long-term light curves on an \textit{INTEGRAL} revolution timescale.  The date of the \textit{NuSTAR} observation is shown with a vertical green line.  The solid black line denotes a flux of 0 mCrab.  The dashed black line denotes the mean flux.} 
\label{fig:lc}
\end{figure}

\section{Results} \label{sec:results}

\subsection{Temporal Variability}

To study the temporal behavior of Cen A, we constructed the SPI and ISGRI light curves on an \textit{INTEGRAL} revolution timescale (\( \sim 2.5 - 3 \) days).  Figure~\ref{fig:lc} shows the temporal variability in the \(20-40\), \(40-60\), \(60-100\), and \(100-200\) keV energy bands.  The SPI data are shown as black diamonds, and the ISGRI data are shown as red triangles.  The time of the \textit{NuSTAR} observation is denoted by a vertical green line.  Both SPI and ISGRI instruments show similar temporal evolutions with significantly higher fluxes above the mean between MJD 54500 and 55000, which are present in the data up to the \(100-200\) keV range.  The source is significantly detected up to the \(200-400\) keV band in SPI at \(11.3 \sigma \) and in the \(200-450\) keV band in ISGRI at \(10.0 \sigma\) during the mission.

\subsection{Spectral Variability}

Numerous authors have reported that Cen A's spectral shape does not change with flux \citep{jourdain1993,rothschild2006,beckmann2011,rothschild2011,burke2014}.  However, recent results from \textit{Swift}/BAT reported spectral softening when the flux increased after 2013 \citep{rani2022}.  Therefore, we searched for spectral variability in the \textit{INTEGRAL} data by fitting the SPI \(25-70\) keV and ISGRI \(34-68\) keV spectra independently to a power-law model.  To reduce uncertainties in the fit parameters, we grouped observations close in time to generate average spectra.  Table~\ref{table:groups} lists the revolution groupings.  Isolated revolutions were excluded from the comparison.

\begin{figure}[h!]
  \begin{center}
  \includegraphics[scale=0.7, angle=180,trim = 155mm 30mm 10mm 0mm, clip]{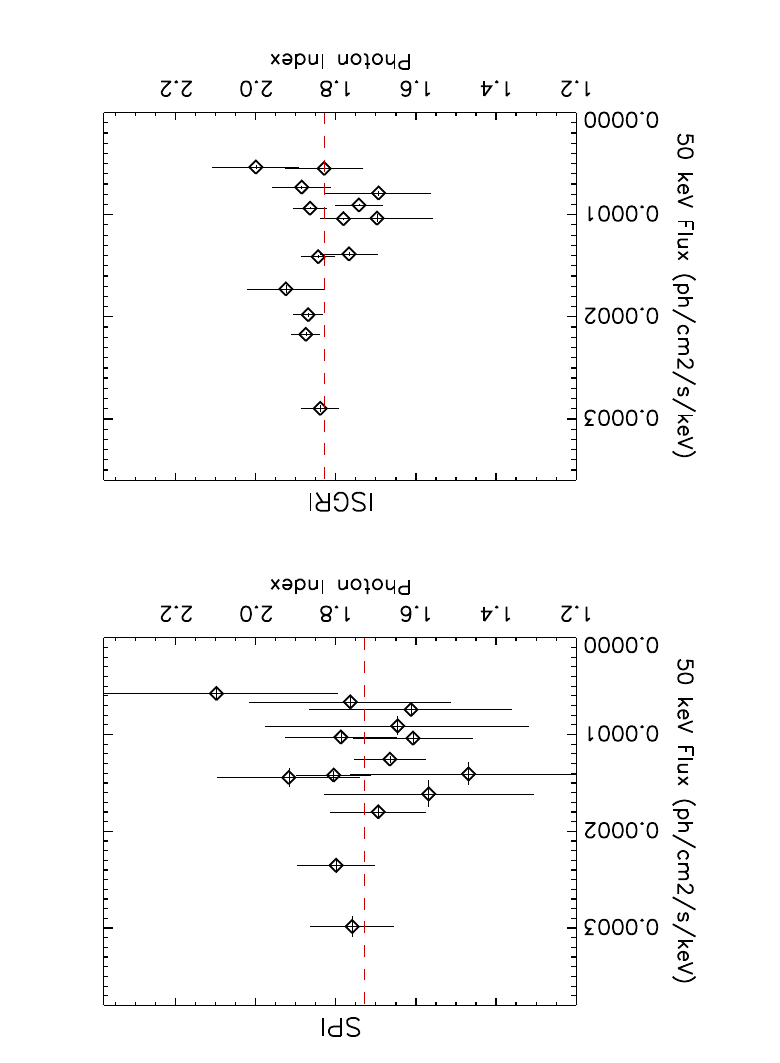}
    \end{center}
  \caption{Plot of photon index versus 50 keV flux for spectral groups for SPI (top) and ISGRI (bottom). The best-fit photon index for each instrument is shown as a red dashed line.} 
\label{fig:spec_compare}
\end{figure}

Figure~\ref{fig:spec_compare} shows the photon index vs 50 keV flux for each grouping in SPI (top) and ISGRI (bottom).  The best-fit photon index values from all the groups were fit to a constant value.  The best-fit average values are \( \Gamma = 1.73 \pm 0.04\) for SPI and \( \Gamma = 1.83 \pm 0.02\) for ISGRI, compared to \(1.75 \pm 0.04\) and \(1.80 \pm 0.01\) for the long-term average spectra of SPI and ISGRI, respectively.  The errors are at the \( 1- \sigma\) level (as are subsequent errors).  The reduced \( \chi^2 / \nu = 0.509 / 13 \) for SPI and \( 1.08 / 13\) for ISGRI.  Therefore, the photon indexes are consistent with a constant value, with the best-fit indexes for both instruments presenting a marginal agreement.  

\begin{wraptable}{r}{4.cm}

\centering
\caption{Spectral Groupings}\label{table:groups}
\vspace{-10mm}
\begin{tabular}{cc} \\\toprule 
\multicolumn{2}{c}{\textit{INTEGRAL}}  \\
\multicolumn{2}{c}{Revolutions}         \\  
\hline
\(48-49\)  &  \(1309-1311\) \\
\(149-163\) & \(1370-1386\)  \\ 
\(514-523\) & \(1437-1439\) \\
\(574-588\) & \(2313-2315\) \\
\(648-651\) & \(2457-2470\)  \\
\(702-705\) & \(2517-2539\) \\
\(764-768\) & \(2581-2590\)  \\
\tableline
\vspace{10mm}
               \end{tabular}
\vspace{20mm}
\end{wraptable}

 We note that the last observing period (rev \(2581-2590\)) showed relatively soft spectra with photon indexes of roughly 2.1 (SPI) and 2.0 (ISGRI), though the errors are large.  During this period, the source was at its lowest flux level observed by \textit{INTEGRAL}, which does not confirm the behavior reported by \cite{rani2022}.

\subsection{Average Spectrum}

\subsubsection{Phenomenological Models}
Since no significant spectral variability was found, the groupings were combined for long-term average spectra to study the high-energy spectrum of Cen A.  Also, we included \textit{NuSTAR} observations, which extends to lower energies, that can test the presence of an Fe line and reflection.  As shown in previous works \citep{burke2014}, a reflection component in the Cen A spectrum can result in larger high-energy cutoff energy.  

The \textit{NuSTAR} data were fit first with an absorbed powerlaw model (\texttt{phabs*phabs*(zpowerlaw+zgauss)}) with absorption components for both the galactic absorption (fixed to \(2.36 \times 10^{20} \) cm\(^{-2}\) \citep{hi4pi2016}) and an intrinsic absorption.  The best parameters from this fit are \( \Gamma = 1.798 \pm 0.002\) with \(n_H = 8.73 \pm 0.07 \times 10^{22}\) cm\(^{-2}\).  The Fe line energy is \(6.36 \pm 0.01\) keV with width of \(0.08 \pm 0.2\) keV with reduced \( \chi^2 = 2581.36/ 2533 = 1.02 \).  The calculated equivalent width from XSPEC is \( \sim 32\) eV.  The best-fit continuum parameters are listed in Table~\ref{table:paras}.  In subsequent fits, the Fe line energy and width were fixed to these values.  

To investigate the presence of a spectral cutoff, we next fit the \textit{NuSTAR} spectra to a \texttt{zcutoffpl} model.  The model better describes the data with a reduced \( \chi^2 = 2469.56/ 2534 = 0.97 \) with a \( \Delta \chi^2 = 111.8 \).  The intrinsic absorption was \(8.0 \pm 0.1 \times 10^{22}\) cm\(^{-2}\).  The fit parameters are given in Table~\ref{table:paras}.

Next we applied the \texttt{zpowerlw} model to the SPI and ISGRI data independently.  The best-fit photon indexes are in perfect agreement between the two instruments for the powerlaw model.  However, neither spectrum was well described by the model, which over-predicted the fluxes above \( \sim 150 \) keV.  The resulting \(\chi^2 / \nu\) values were \(63.18/34\) for SPI and \(64.20/ 43\) for ISGRI.  

Due to the implied presence of curvature in the spectra, we next fit the data with a cutoff power-law model (\texttt{zcutoffpl}).  The \(\chi^2 / \nu\) values improved to \(31.15/33\) and \(38.66/42\) for SPI and ISGRI, respectively.  The best-fit parameters in Table~\ref{table:paras} are in agreement between the two instruments.  Thus we performed a joint fit with the two instruments.  A cross-instrument normalization was added to the model with the SPI normalization fixed.  The model was still an acceptable fit to the data with a \(\chi^2 / \nu\) of \(77.81/79\).  The cross-instrument normalization was \(0.873 \pm 0.008\).  

Subsequently, we performed a joint fit between the \texttt{NuSTAR}, SPI, and ISGRI data covering \(3.5-2200\) keV.  A power-law fit to the data is statistically acceptable with a reduced \( \chi^2 = 2795.87/ 2610 = 1.07 \), but it is unsurprisingly a worse fit due to the high-energy cutoffs seen in the three instruments.  As expected, including a high-energy cutoff improves the quality of the fit (reduced \( \chi^2 = 2625.40/ 2609 = 1.01 \)), but with a relatively high cutoff energy of \( \sim 650\) keV.  The cross-normalization values for the \textit{NuSTAR} spectra were \(1.81 \pm 0.02\) for FPMA and \(1.80 \pm 0.02\) for FPMB.  Subsequent spectral fits result in similar cross-normalization values.  These values are large compared to the long-term average SPI spectrum because the \textit{NuSTAR} observation occurs while Cen A is in a high-flux state.  

To investigate the presence of a reflection component, which can result in artificially high cutoff energies \citep{burke2014}, we used the \texttt{pexrav} continuum model.  The best-fit reflection fraction was \( 2 \times 10^{-7} \pm 1 \times 10^{-2}\).  We also tested the presence of reflection with relativistic smearing using a \texttt{xion*(pexrav+zgauss)} model.  We assumed a geometry with a central hot sphere and a cold outer disk.  The model describes the data well (red. \( \chi^2 = 2503.97 / 2527 = 0.991\)), but most of the fit parameters are unconstrained, including the reflection fraction (\(0.0 \pm 0.2\)).  Thus we consider that no significant reflection is required by the data.  Consequently, the reflection fraction was fixed to 0.  The spectrum was refit, and the best-fit parameters (Table~\ref{table:paras} and subsequent fit parameters; \(\chi^2 / \nu = 2594.66/2608 = 0.99\)) are the same with those of the \texttt{zcutoffpl} model.  

\begin{table*}
\begin{center}
\caption{Average Spectrum Parameters}
 \hspace{-30mm}
\scalebox{0.85}{
\hspace{-6mm}
\begin{tabular}{ccccccccc}
\tableline \\
\multicolumn{1}{c}{} & \multicolumn{3}{c}{\texttt{ZPowerlw}}  & \multicolumn{4}{c}{\texttt{Cutoff ZPowerlaw}} &  \\
\tableline
Inst.            & \(n_H\)                              &   \( \Gamma\)         & \(\chi^2 / \nu\)       & \(n_H\)                                   &   \(\Gamma\)        & \(E_{cut}\)     & \(\chi^2 / \nu\)   \\
                 &  (\(\times 10^{22}\) cm\(^{-2}\))    &                       &                        &  (\(\times 10^{22}\) cm\(^{-2}\))         &                     & (keV)           &                     \\
\tableline                   \\
\textit{NuSTAR}\(^a\)& \(8.73 \pm 0.07\)&  \(1.798 \pm 0.002\) & 1.02  & \(8.0 \pm 0.1\)         &  \(1.725 \pm 0.007\)& \(246 \pm 24\) & 0.97    \\
SPI             & \(-\)            & \(1.86 \pm 0.02\)   & \(1.86\) & \(-\)      &  \(1.62 \pm 0.05\)  & \(381 \pm 85\) &  0.94         \\ 
ISGRI           & \(-\)            &\(1.851 \pm 0.007\) & \(1.49\)  & \(-\)    &  \(1.72 \pm 0.03\)  & \(517 \pm 112\) &  0.92        \\ 
SPI/ISGRI       & \(-\)            &\(1.859 \pm 0.007\) & \(1.51\)   &\(-\)     &  \(1.72 \pm 0.02\)  & \(543 \pm 94\) &  0.98         \\
\textit{NuSTAR}/SPI/ISGRI\footnote[1]{\texttt{zgauss} model included in fit with fixed parameters to model Fe line.  (See Sec. 3.3.1.)} &\(8.94 \pm 0.07\)& \(1.807 \pm 0.002\)  & 1.071 & \(8.37 \pm 0.08\) & \(1.767 \pm 0.004\) & \(638 \pm 54\) & 1.00         \\ 

\tableline
\end{tabular}
}

\label{table:paras}

\end{center}
\end{table*}

\subsubsection{Physically Motivated Models}

To better understand the emission mechanism(s) in this energy range, we fit the data using more physically motivated models.  We fit the data to a \texttt{CompTT} model. As pointed out in \cite{furst2016}, the seed photon temperature is unknown thus they fit their data to a range of seed photon temperatures with extremes of \(0.05\) keV and \(0.5 \) keV.  Therefore, we fit our data to minimum and maximum values to explore the range of electron temperatures.  Both fits find comparable values of \( kT_e \sim 520 \) keV and \( \tau \sim 0.03\).  The data prefer the lower \( kT_0 \) (0.05 keV) with \( \Delta \chi^2 = 21.64\), though both models acceptably describe the data.  We also tried a fit with \(kT_0\) free, finding a best-fit seed photon temperature of \(0.4 \pm 0.1\) keV and a reduced \(\chi^2 =2588.54 /2607 = 0.99 \), \(\Delta \chi^2 =5.34\) for one additional free parameter compared to the fit with \( kT_0 = 0.05\) keV.  The best-fit parameters for each fit are listed in Table~\ref{table:joint_spec}.  We note that implied electron temperatures above 500 keV disfavor a thermal origin \citep{fabian2017}.

\begin{figure}[h!]
  \centering
  \includegraphics[scale=0.7, angle=0,trim = 15mm 125mm 0mm 40mm, clip]{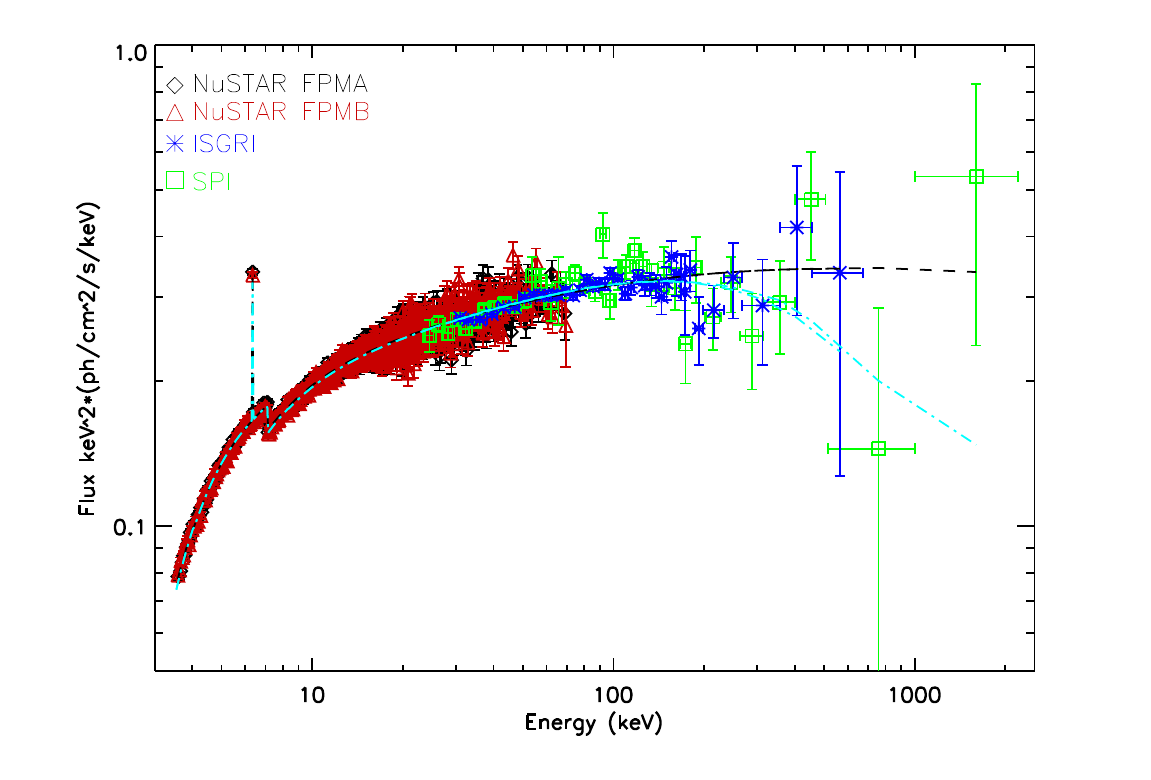}
  \caption{Cen A long-term average spectra from \textit{INTEGRAL} SPI (green square) and ISGRI (blue stars) with \textit{NuSTAR}/FPMA (black diamonds) and FPMB (red triangles).  The best-fit \texttt{CompTT} with \(kT_e = 0.5\) keV model (cyan dot-dash line) and log parabola model (black dashed line) are overplotted.} 
\label{fig:spectra}
\end{figure}

As discussed above, the hard X-ray/soft gamma-ray emission is potentially due to SSC jet emission.  Thus, we fit the spectra to a log parabola model, which can model the inverse Comptonization found in \textit{Fermi} blazars (See \cite{esposito2015} and references within).  The \( \alpha\) value in the \texttt{log-parabola} is the slope observed at the pivot energy (\(F(E) = N (E/E_{pivot})^{-\alpha -\beta log (E/E_{pivot})}\)).  Thus, we fixed the \(E_{pivot}\) to 50 keV to compare this slope with previously reported power-law results. This model also provides a good fit to the data (reduced \( \chi^2 = 2553.05/2608  = 0.98 \)) with \( \Delta \chi^2 = 41.45 \) for the same number of free parameters compared to the \texttt{pexrav} model and \( \Delta \chi^2 = 35.49 \) for 1 fewer free parameter compared to the \texttt{CompTT} model with \(kT_0\) free.  Figure~\ref{fig:spectra} shows the \textit{NuSTAR} spectrum in black diamonds (FPMA) and red triangles (FPMB), SPI spectrum in green squares, and the ISGRI spectrum in blue stars with the \texttt{CompTT} with \(kT_e = 0.5\) keV (cyan dot-dashed line) and log parabola (black dashed line) models.  Therefore, we find that statistically both thermal Comptonization and SSC emission can acceptably describe the joint \textit{NuSTAR}/\textit{INTEGRAL} spectrum.  Below, we discuss the implications of the parameters to potentially differentiate the two origins.

\begin{table*}
\begin{center}
\caption{Joint \textit{NuSTAR/SPI/ISGRI} Fit Spectral Parameters}
 \hspace{-30mm}
\scalebox{0.85}{
\hspace{-6mm}
\begin{tabular}{cccccccccccc}
\tableline \\
 \multicolumn{5}{c}{\texttt{CompTT}}  & \multicolumn{4}{c}{\texttt{Log Parabola}} \\
\tableline
\(N_H\) & \(kT_0\)   &    \( kT_e\)        & \( \tau \) & \(\chi^2 / \nu\)  & \(N_H\) & \(\Gamma \) &  \(b\) & \(\chi^2 / \nu\) \\
 (\(\times 10^{22}\) cm\(^{-2}\))  &   (keV)   &       (keV)        &            &                   &             &                            \\
\tableline                   \\
\(8.38 \pm 0.08\) & 0.05 (fixed)            & \(535 \pm 53\)     & \(0.029 \pm 0.006\)  & 1.01          & \(7.8 \pm 0.1\) & \(1.830\pm 0.003\) &\(0.069 \pm 0.005\)& \(0.98\)\\ 
\(8.66 \pm 0.08\) & 0.5  (fixed)            & \(525 \pm 53\)     & \(0.029 \pm 0.006\)  & 0.99          &                       &        & \\ 
\(8.5 \pm 0.2\) & \(0.4 \pm 0.1\)  & \(556 \pm 62\)     & \(0.026 \pm 0.006\)  & 0.99            &                       &        & \\
\tableline
\end{tabular}
}

\label{table:joint_spec}
\end{center}
\end{table*}

\section{Discussion} \label{sec:disc}

\subsection{Implications of Model Parameters}

\subsubsection{Iron Line and Reflection}

A Fe K\( \alpha\) line is suggestive of the presence of photons from an X-ray source reflecting off of a surrounding medium \citep{george1991}.  The width of the line profile provides an indication of the distance between the source and the reflecting medium \citep{evans2004}.  The width of \( \leq 8 \) eV found in this work is similar to \cite{furst2016} indicating that the reflecting medium is far from the central black hole.  

Additionally, the column density and the equivalent width of the line are related to the geometry and elemental abundances of the material.  The calculated equivalent width from XSPEC is \( \sim 32\) eV.  This value combined with the fit column densities (\( \sim 7-9 \times 10^{22}\) cm\(^{-2}\)) is compatible with a face-on torus interpretation with no significant Compton hump expected \citep{matt2003}, which is consistent with the reflection value from our \texttt{pexrav} fit.  Such an interpretation is in agreement with MYtorus and BNtorus fit results in \cite{furst2016} and infrared observations from \cite{ramos2009}.  Absence of reflection has also been reported in other radio-loud AGN \citep{sambruna1999,eracleous2000,reves2000,grandi2001,ballantyne2002,ogle2005,ballantyne2007,sambruna2009,evans2010,walton2013,ballantyne2014,madsen2015}.  There are other potential explanations for the lack of Compton hump: a truncated inner disk from a retrograde spin \citep{garofalo2009}, an ionized accretion disk \citep{ballantyne2002}, an outflowing corona \citep{beloborodov1999,malzac2001}, relativistic jet dilution \citep{reves2000}, changing geometry of the inner disk \citep{eracleous2000}, or central accretion flow obscuration by the jet \citep{sambruna2009}.  All of these models are consistent compared to result in Cen A.

\subsubsection{Continuum}

When including data up to the soft gamma-rays, we find spectral curvature is present in Cen A.  Even the cutoff power-law fit to the \textit{NuSTAR} spectrum is preferred over the power-law fit.  This contrasts with spectral fits based on hard X-ray instruments that reported no spectral cutoff until the MeV region \citep{rothschild2011,furst2016}.  The cutoff energy value varies across the instruments (\(\sim 250 - 520\) keV) in this work.  This behavior is similar to what has been reported in other works using soft gamma-ray data, since the parameter is sensitive the energy domain of the instruments.  

Our joint \textit{NuSTAR}/SPI/ISGRI fit has a best-fit cutoff energy of roughly 650 keV or an electron temperature of \( \sim 520-550\) keV.  This implies a corona near the runaway pair-production zone of the compactness-temperature diagram for a normal corona size of \(3-10\) \(r_g\) \citep{fabian2015,fabian2017}.  These results together with the lack of significant reflection suggest that the continuum in the hard X-ray is due to SSC emission.  However, we are not able to exclude coronal emission as the dominant emission process in the hard X-ray/soft gamma-ray band.  For example, 3C 273 showed similar Fe line and reflection characteristics and analysis of the hard X-ray/soft gamma-ray spectrum continuum was found to be a combination of coronal plus jet emission \citep{grandi2004,madsen2015}.  

\subsection{Joint Fit With Fermi/LAT}

Following \cite{madsen2015}, we included \textit{Fermi}/LAT data using the 12-year Data Release 3 catalog fluxes in 8 channels over the 50 MeV \( -\) 1 TeV energy range \citep{abdollahi2022}.  Joint LAT/HESS analysis found a spectral break at \( \sim 3\) GeV \citep{hess2018}.  Thus, we included a power-law model to fit the data above 3 GeV for a continuum model of \texttt{zpowerlw} in fits to the LAT data.

\subsubsection{Jet Only Scenario}

Initially, we tested a model where the keV to GeV emission is due to the jet (\texttt{log-parabola + zpowerlw}) with the LAT cross-normalization parameter (\(C_{LAT}\)) free and fixed to 1.  The parameters for this fit are \(n_H = 8.1 \pm 0.1 \times 10^{22}\) cm\(^{-2}\), \( \alpha = 1.83 \pm 0.009 \) (\(E_{pivot} = 50\) keV), \(\beta = 0.115 \pm 0.008 \), and \(\Gamma_{HE} = 2.17 \pm 0.02\) with reduced \( \chi^2 = 2490.84/2613 = 0.99 \) and \(C_{LAT} =  0.38 \pm 0.09\).  When the cross-normalization is fixed to 1, the parameters are \(N_H = 8.3 \pm 0.1 \times 10^{22}\) cm\(^{-2}\), \( \alpha = 1.826 \pm 0.007 \) (\(E_{pivot} = 50\) keV), \(\beta = 0.156 \pm 0.003 \), and \(\Gamma_{HE} = 2.250 \pm 0.005\) with reduced \( \chi^2 = 2611.81/2614 = 1.00 \).  We can describe the data with both \(C_{LAT}\) values, but the free value implies a LAT flux significantly higher than the flux (\(\times \sim 2.6\)) reported by the LAT team \citep{abdollahi2022}.  Thus we fixed \(C_{LAT}\) to 1 in subsequent fits.  

\subsubsection{Jet and Corona Scenario}

Also, we tested a model where the coronal emission dominates the hard X-ray/soft gamma-ray emission and the jet component dominates in the MeV to GeV range, similar to the scenario presented for 3C 273 in \cite{grandi2004} and \cite{madsen2015}.  We used a (\texttt{CompTT + log-parabola + zpowerlw}) continuum model.  However, some parameters are poorly constrained due to the lack of data in the MeV range and the GeV power law component.

We found \(n_H = 8.40 \pm 0.09 \times 10^{22}\) cm\(^{-2}\), \(kT_e = 481 \pm 121\) keV, \( \tau = 0.03 \pm 0.01\), \( \alpha = 2.6 \pm 0.1 \) at an \(E_{pivot} = 1 \times 10^5 \) keV, \(\beta = 0.3 \pm 0.3 \), and \(\Gamma_{HE} = 2.1 \pm 0.2\) with reduced \( \chi^2 = 2611.60/2612 = 1.00 \). We report only the results with \(kT_0\) fixed to 0.5 keV as the fit to \(kT_0=0.05\) keV results in similar poorly constrained values.  We obtain a \(kT_e\) near the runaway limit, but with a large error.

The fit well describes the data, but the high-energy power-law over-predicts the observed high-energy flux seen by LAT and H.~E.~S.~S.  Thus, we fixed the photon index value found in the jet only scenario (2.25) and refit the spectra.  The new model results in comparable parameters without much change in constraints and in goodness of fitwhen including a \texttt{expabs} to the \texttt{zpowerlw} to limit its low-energy contribution. \( \chi^2 = 2623.95/2613 = 1.00 \).  The low-energy cutoff energy was fixed to \(10^5\) keV.

\begin{figure}[h!]
  \centering
  \includegraphics[scale=0.5, angle=180,trim = 15mm 15mm 0mm 20mm, clip]{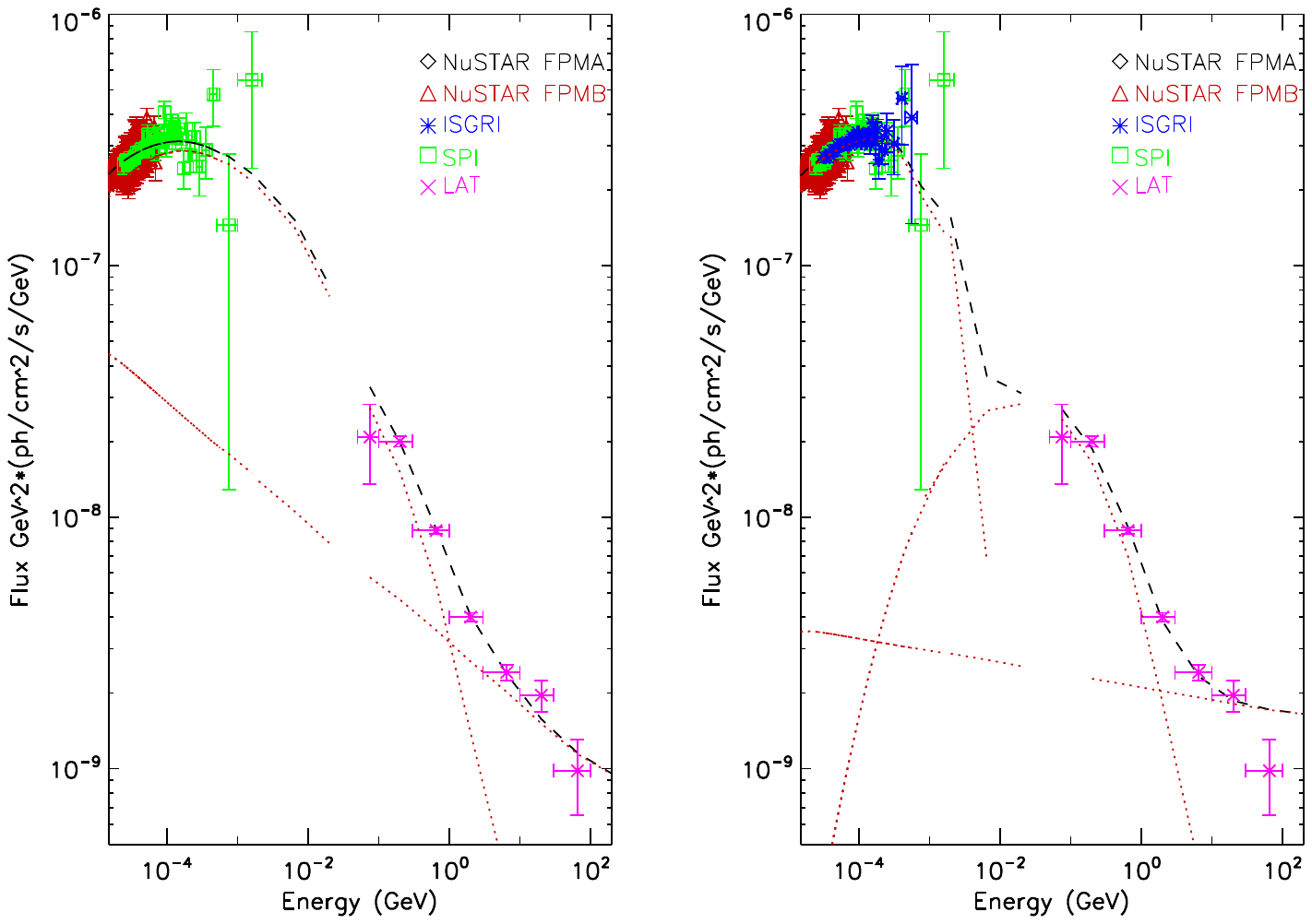}
  \caption{Cen A long-term average spectra from \textit{INTEGRAL} SPI (green square), ISGRI (blue stars) with \textit{NuSTAR}/FPMA (black diamonds) and FPMB (red triangles) and LAT (magenta x's).  The left panel shows the \texttt{logparabola+zpowerlw} model (black dashed line) with the individual continuum components shown as red dotted lines.  The right panel shows the \texttt{CompTT+logparabola+zpowerlw} model  (black dashed line) with individual continuum components as red dotted lines.} 
\label{fig:spectra_lat}
\end{figure}

Figure~\ref{fig:spectra_lat} shows the \textit{NuSTAR} spectrum in black diamonds (FPMA) and red triangles (FPMB), SPI spectrum in green squares, the ISGRI spectrum in blue stars, and LAT spectrum in magenta x's with both \texttt{logparabola+zpowerlw} (left) and \texttt{CompTT+logparabola+zpowerlw} (right; where the hard X-rays are due to the corona and the MeV to GeV flux is from the jet) models.  The differences between the models can be seen in the MeV region where the jet only scenario smoothly connects from hard X-rays to high-energy gamma-rays.  In contrast, the other scenario predicts a drop in flux above a few MeV.  Thus, using the long-term average spectra from \textit{INTEGRAL}/SPI and ISGRI and LAT, we are not able to differentiate between the possibilities without observations in the medium MeV region.

Note that \cite{rani2022} proposed a scenario where the peak energy of the jet overlaps with the corona component at \(\sim 500\) keV, thus hiding the corona component. However, we were not able to disentangle the presence of the two components with our data.

\subsubsection{Comparison with Previous Works}

In their work on Cen A, \cite{rani2022} also showed that the 230 GHz flux correlates with the hard X-ray flux with no time lag.  (\cite{israel2008} found similar results.)  In addition, they found the 230 GHz flux is higher than what is predicted for an advection dominated accretion flow (ADAF), and the radio spectrum is inconsistent with the slope predicted for an ADAF.  Thus the authors conclude a jet component is enough to explain the hard X-ray emission.  Our work supports that conclusion including X-ray to GeV data.

The source 3C 273 provides an interesting comparison since it was the first AGN where both corona and jet components have been reported \citep{grandi2004}. However, the Cen A jet component peaks at \( \sim 50\) MeV compared to roughly 2 MeV for 3C 273.  Additionally, for Cen A, the SED peak is dominated by the corona emission (around 500 keV) while the jet component dominates the SED peak in 3C 273 \citep{madsen2015}. Therefore, it seems probable that even if a corona exists and contributes to the hard X-ray emission in Cen A, it would be hard to observe and distinguish from the jet emission.

\section{Conclusion} \label{sec:con}

In this work, we analyzed \textit{INTEGRAL}/ISGRI and SPI observations of Cen A covering roughly 20 years.  We first searched for spectral variability on 
a timescale of \(\sim \) months, and did not find any significant deviations from a power-law with a constant value.  Building from this result, we constructed a long-term average spectrum up to 2.2 MeV. We included an observation from \textit{NuSTAR}, which provides coverage down to 3.5 keV, to study the X-ray to soft gamma-ray spectrum.  

Analysis of this spectrum found a narrow Fe line (\(< 8 \) eV) and no significant reflection component, which are consistent with previous results (apart from \cite{burke2014} that report marginal reflection).  The joint \textit{NuSTAR}/ISGRI/SPI spectral fit has a cutoff energy of \( \sim 650\) keV or an electron temperature of \(\sim 520 - 550\) keV, depending on the seed temperature.  The physical interpretation of the hard X-ray/soft gamma-ray emission with these values is ambiguous, as the \(kT_e\)'s are near the runaway pair-production region for the corona.  Another possible scenario is synchrotron self-Compton emission from the jet, which can also well describe the spectrum using a log-parabola model.  

To complete our overview of the source, the LAT spectrum in the 50 MeV to 100 GeV range was included to the spectral fit.  A fit to a \texttt{log-parabola} model for the jet emission explains the spectrum from \textit{NuSTAR} to LAT below \( \sim 3\) GeV.  However, the joint spectrum can also be adequately described by the inclusion of a \texttt{CompTT} component to investigate the presence of coronal emission.  

The results with this component are degenerate, however, but the best-fit parameters indicate two general scenarios.  One where the X-ray emission is completely due to the corona and the jet emission dominates at MeV/GeV energies.  The other is that the jet flux can explain the spectrum from X-rays to the GeV region.  Our work favors the jet interpretation as one component can explain the whole emission from keV to LAT and due to the high electron temperature implied by the corona scenario.  This interpretation is consistent with the polarization results by IXPE, which did not detect significant X-ray polarization \citep{ehlert2022}.  In their scenario, the X-rays can be explained by SSC emission with electrons accelerated in magnetic fields that are highly disorganized around the innermost jet.


To definitively exclude/constrain the presence of a corona component, several possibilities can be explored.  As discussed above, the high-energy gamma-ray spectrum has two components, which limits the contraints on the curvature of the jet spectrum from LAT observations that can be extrapolated to lower energies.  A detailed analysis of the Cen A long-term spectrum in high-energy gamma-rays (potentially including H.~E.~S.~S. data) may help to better constrain the hard X-ray/soft gamma-ray emission.  

Another constraint can come from a significant Cen A detection at energies between SPI and LAT to differentiate between the jet only model and the corona \(+\) jet scenario as the predicted fluxes are significantly different in the MeV range (Figure~\ref{fig:spectra_lat}).  Future observations by COSI may provide clarification, but the energy range is expected to extend up to only 5 MeV \citep{tomsick2019}.  Thus observations from a proposed mission such a AMEGO-X will be the best tool to study the full MeV energy range \citep{caputo2022}.  

\begin{acknowledgments}

We thank the referee for their comments and input.  The research leading to these results has received funding from the European Union’s Horizon 2020 Programme under the AHEAD2020 project (grant agreement n. 871158).  J.R. acknowledges financial support under the INTEGRAL ASI/INAF No. 2019-35.HH.0 and from an INAF 2022 mini-grant.  M.M. acknowledges financial support from ASI under contract n. 2019-35-HH.0.  The \textit{INTEGRAL} SPI project has been completed under the responsibility and leadership of CNES.  We are grateful to ASI, CEA, CNES, DLR, ESA, INTA, NASA and OSTC for support.


\end{acknowledgments}

\end{document}